\documentclass[aps,a4paper,showpacs,amsfonts,amsmath,amssymb]{revtex4}

\usepackage{graphicx}
\usepackage{pstricks}
\usepackage{amsmath}
\newpsobject{showgrid}{psgrid}{subgriddiv=1,griddots=10,gridlabels=6pt}

\begin{document}
\title{From Classical to Quantum Mechanics through Optics}

\author{Jaume Masoliver$^1$, Ana Ros$^1$}
\address{$^1$ Departament de F\'{\i}sica Fonamental. Universitat de Barcelona.\\Diagonal 647, 08028 Barcelona, Spain.}

\begin{abstract}
In this paper we revise the main aspects of the Hamiltonian analogy: the fact that optical paths are completely analogous to mechanical trajectories. We follow Schr\"{o}dinger's original idea and go beyond this analogy by changing over from the Hamilton's principal function $S$ to the wave function $\Psi$. We thus travel from classical to quantum mechanics through optics.
\end{abstract}
\pacs{0155, 4000, 0300, 0165}
\maketitle

\section{Introduction}

According to Arnold Sommerfeld mechanics is ``the backbone of mathematical physics''; it is a highly elaborated theory emerging from Newton's laws which, in turn, summarize in precise form the whole body of experience. 

Classical mechanics describes with great precision the motion of macroscopic bodies, including astronomical bodies, except when velocities and masses involved are excessively large. In the domain of microscopic bodies: atoms, molecules, electrons, etc., the situation is quite different, for the laws of mechanics, as well as those of electromagnetism, do not explain satisfactorily the experimental observations, and are not appropriate for describing motion at a microscopic scale. During the first third of the 20th century, a new mechanics was developed --named ``quantum mechanics'' by Max Born in 1924-- which governs the laws of motion for microscopic bodies but that coincides with classical mechanics when applied to macroscopic bodies.

Quantum mechanics is often presented as a brand-new development based on axioms, some of them counterintuitive, and  which seem to be hardly related to classical mechanics and other precedent theories. The axiomatic approach has indeed methodological advantages, but overlooks the fact that quantum theory is intimately related with classical mechanics, specially with its Hamiltonian formulation. Bohr's theory of electronic orbits employed Hamiltonian methods when it was realized the importance of separable systems in the formulation of the quantum conditions of Sommerfeld and Wilson in 1916; and also in the   calculations of the Stark effect performed by Epstein the same year. 

The reinterpretation of the quantum laws given by Schr\"{o}dinger, Heisenberg and Dirac also emanated from Hamiltonian methods. The matrix nature of the canonical variables $(q,p)$ was introduced by Heisenberg, Born and Jordan; while Dirac considered the conjugated variables as non-commutative operators. On the other hand, Schr\"{o}dinger developed the operational point of view, and going farther than the analogy between optics and mechanics already established by Hamilton, turned the classical Hamilton-Jacobi equation into a wave  equation. There is, therefore, a passage going from classical mechanics to quantum mechanics through optics, a path taken by Schr\"{o}dinger one century after Hamilton.

In 1831 William Rowan Hamilton imagined the analogy between the trajectory of material particles moving in potential fields and the path of light rays in media with continuously variable refractive indices. Because of its great mathematical beauty, the ``Hamiltonian analogy'' survived in textbooks of dynamics for almost a hundred years, but did not stimulate any practical application until 1925, when H. Busch explained the focusing effect of the  electromagnetic field on electron beams in optical terms, which would inspire the quick development of electronic microscopy from 1928 onwards. Almost simultaneously, in 1926, Erwin Schr\"{o}dinger went one step beyond and, using the ideas presented by Louis de Broglie in 1924, moved from geometric optics to wave optics of particles . In the present work we remember and reconstruct this passage.

\section{Classical mechanics}

A typical mechanical system, to which many physical systems are reduced, consists in a collection of mass points interacting between them following well-known laws. Experience shows that the state of the system is entirely determined by the assembly of positions and velocities of all its particles. The reference system chosen to describe positions and velocities needs not to be Cartesian: the representation of the system can be achieved by means of generalized coordinates, $\mathbf{q}=(q_1,q_2,\cdots,q_n)$, and generalized velocities, $\mathbf{\dot{q}}=(\dot{q}_1,\dot{q}_2,\cdots,\dot{q}_n)$. The minimum number, $n$, of generalized coordinates which are needed to describe completely the state of the system is called the number of degrees of freedom.

The laws of mechanics are those that determine the motion of the system by providing the time evolution of the positions, $\mathbf{q}(t)$, and the velocities , $\mathbf{\dot{q}}(t)$, of all its components. Perhaps the most general and concise way of stating these laws is through a variational principle known as the {principle of least action} or {Hamilton's principle}:
\begin{equation}
\delta\left(\int_{t_1}^{t_2} L dt\right)=0,
\label{pma}
\end{equation}
where the symbol $\delta$ stands for an infinitesimal variation of the trajectory. Here the  Lagrangian,  $L=L(\mathbf{q},\mathbf{\dot{q}},t)$, is a function of the generalized coordinates and velocities of every particle of the system. It is a general characteristic of the equations of classical physics that they can be derived from variational principles. Some examples are Fermat's principle in optics (1657) and Maupertuis'principle in mechanics (1744). Equations from elasticity, hydrodynamics and electrodynamics can also be derived in the same way.

The variational principle expressed by (\ref{pma}) means that when the system moves from some initial configuration at a given time $ t_1$ to a final configuration at some other time $t_2$, the real trajectory is such that the action integral (\ref{pma}) reaches  a stationary value; it does not matter whether it is a minimum or a maximum or a saddle point. The main advantage of this formulation is its independence of the system of coordinates, a property that Newton's equations of motion do not enjoy.

The variational principle (\ref{pma}) leads to the {Lagrange equations}:

\begin{equation}
\frac{d}{dt}\frac{\partial L}{\partial\mathbf{\dot{q}}}-
\frac{\partial L}{\partial\mathbf{q}}=0,
\label{lagrange}
\end{equation}
where we have used the notation
$$
\frac{\partial L}{\partial\mathbf{q}}=\left(\frac{\partial L}{\partial q_1},\frac{\partial L}{\partial q_2},\cdots,\frac{\partial L}{\partial q_n}\right)
$$
for the gradient of $L$ with respect to $\mathbf{q}$, and similarly for $\partial L/\partial\mathbf{\dot{q}}$. 

Once we know $L$, Lagrange equations constitute a set of $n$ second-order differential equations in time which, along with fixed initial conditions, determine the time evolution of the system. For a system of particles without electromagnetic interactions and with masses and velocities not too large, so that relativistic effects can be neglected, the Lagrangian is determined in such a way that the set of differential equations (\ref{lagrange}) coincides with Newton's equations of motion:
$$
\frac{d\mathbf{p}}{dt}=-\frac{\partial V}{\partial\mathbf{q}},
$$
where $V=V(\mathbf{q})$ is the potential energy of the system. This can be achieved if 
$$
L=T-V,
$$
where $T$ is the kinetic energy.

In the Lagrangian formulation the state of the system is governed by a set of second-order differential equations, each one involving second-order derivatives of the coordinates  with respect to time. In many cases, specially in general theoretical derivations such as the present one, it is convenient to replace Lagrange equations by an equivalent set of twice as many equations of first-order. The most  direct way of accomplishing this would be setting  $\mathbf{\dot{q}}=\mathbf{v}$ and then add these additional equations to the set (\ref{lagrange}), treating $\mathbf{q}$ and $\mathbf{v}$ as unknowns. 

However, a much more symmetrical formulation is obtained as follows; instead of the velocities, we introduce the new variables:
\begin{equation}
\mathbf{p}=\frac{\partial L}{\partial\mathbf{\dot{q}}},
\label{momenta}
\end{equation}
called {\it momenta}. Substituting this into (\ref{lagrange}) we have
\begin{equation}
\mathbf{\dot{p}}=\frac{\partial L}{\partial\mathbf{q}}.
\label{dotp}
\end{equation}
Therefore, we have reduced the system of $n$ second-order equations given by (\ref{lagrange}) to a system of $2n$ first-order equations given by (\ref{momenta}) and ({\ref{dotp}). However, the latter system is not symmetrical because velocities appear mixed with positions and momenta. The next step is to introduce, instead of the Lagrangian $L(\mathbf{q},\mathbf{\dot{q}},t)$, a new function $H(\mathbf{q},\mathbf{p},t)$ through a Legendre transformation
\begin{equation}
H=\mathbf{\dot{q}}\cdot\mathbf{p}-L,
\label{hamiltonia}
\end{equation}
where we use the notation $\mathbf{\dot{q}}\cdot\mathbf{p}=\sum_{k}\dot{q}_kp_k$. Using $H$, it is not difficult to show that Lagrange equations take the symmetrical form:
\begin{equation}
\dot{\mathbf{q}}=\frac{\partial H}{\partial\mathbf{p}},\qquad
\dot{\mathbf{p}}=-\frac{\partial H}{\partial\mathbf{q}},
\label{hamilton}
\end{equation}
which is the so called canonical form of the equations of motion or {Hamilton equations}. Combining the definition of $H$ given by (\ref{hamilton}) with the Lagrange equations (\ref{lagrange}), we also find that
$$
\frac{dH}{dt}=\frac{\partial L}{\partial t}.
$$
Thus, if $L$ does not depend explicitly of time, then  $\partial L/\partial t=0$ and consequently $H=$ constant. In this case, we say that the system is conservative.

The function $H(\mathbf{q},\mathbf{p},t)$ is called the {Hamiltonian} of the system. In Newtonian mechanics, where $L=T-V$ and $T$ is a quadratic and homogeneous function of the velocities, we can see from (\ref{hamilton}) that $H=T+V$ and the Hamiltonian coincides with the energy. Furthermore, if the system is conservative, then 
$H=E=$ constant, which is the law of conservation of energy. We should note that this discussion is only valid for inertial reference frames and not for accelerated ones. For non-inertial systems as, for instance, a rotating reference frame, the Hamiltonian H is constant but does not coincide with the energy.

\section{The Hamilton-Jacobi equation}

For a mechanical system with $n$ degrees of freedom, Hamilton equations are a set of $2n$ first-order differential equations which must be solved with given initial conditions. The state of the system is specified by coordinates and momenta instead of coordinates and velocities of the Lagrangian formulation. Therefore, knowing the state of the system at a given time, the solution of the differential equations (\ref{hamilton}) will determine the state of the system at any subsequent time.
	
Within the Hamiltonian framework, there exists an alternative way of finding the time evolution of the system. This alternative is based on the theory of {canonical transformations} (i.e., changes of variables leaving invariant the form of Hamilton equations). It relies on the knowledge of certain function, $S(\mathbf{q},t)$, depending on the generalized coordinates and time, called {Hamilton's principal function}. Contrary to the Lagrangian and Hamiltonian approaches, where many ordinary differential equations have to be simultaneously solved, Hamilton's principal function satisfies one single equation, a non-linear partial differential equation of first order called the {Hamilton-Jacobi equation}:
\begin{equation}
\frac{\partial S}{\partial t}+
H\left(\mathbf{q},\frac{\partial S}{\partial\mathbf{q}},t\right)=0,
\label{hj}
\end{equation}
where $H$ is the Hamiltonian in which we have replaced momenta $\mathbf{p}$ by the gradient of S, $\boldsymbol{\nabla} S=\partial S/\partial\mathbf{q}$.

The {complete solution} of the Hamilton-Jacobi equation is any solution depending on $n$ arbitrary and independent constants $\boldsymbol{\alpha}=(\alpha_1,\alpha_2,\cdots,\alpha_n).$ 
It can be shown that within this formalism \cite{born,goldstein}, the time evolution of the system is given by the complete solution $S(\mathbf{q},\boldsymbol{\alpha},t)$ through its derivatives:
\begin{equation}
\mathbf{p}=\frac{\partial}{\partial\mathbf{q}} S(\mathbf{q},\boldsymbol{\alpha},t),
\qquad \boldsymbol{\beta}=
\frac{\partial}{\partial\boldsymbol{\alpha}} S(\mathbf{q},\boldsymbol{\alpha},t),
\label{solucio_hj}
\end{equation}
where $\boldsymbol{\beta}=(\beta_1,\beta_2,\cdots,\beta_n)$ are another set of $n$ arbitrary constants. Note that equations (\ref{solucio_hj}) determine, in an implicit way, the equations of motion 
$$
\mathbf{q}=\mathbf{q}(t;\boldsymbol{\alpha},\boldsymbol{\beta}) \qquad\qquad \mathbf{p}=\mathbf{p}(t;\boldsymbol{\alpha},\boldsymbol{\beta}),
$$
which depend on 2n arbitrary constants $\boldsymbol{\alpha}$ and  $\boldsymbol{\beta}$.

In the case of a conservative system where the Hamiltonian does not depend explicitly on time and is a constant of motion, $H(\mathbf{q},\mathbf{p})=E$, the time dependence of the complete solution of the Hamilton-Jacobi equation is very simple and reads
\begin{equation}
S(\mathbf{q},\boldsymbol{\alpha},t)=-Et+W(\mathbf{q},\boldsymbol{\alpha}),
\label{solucio_hjc}
\end{equation}
where $W(\mathbf{q},\boldsymbol{\alpha})$ a time independent function called {\it Hamilton's characteristic function}.  Substituting (\ref{solucio_hjc}) into (\ref{hj}) we see that this function satisfies the following differential equation, also called Hamilton-Jacobi equation,
\begin{equation}
H\left(\mathbf{q},\frac{\partial W}{\partial\mathbf{q}}\right)=E,
\label{hjc}
\end{equation}
where the dependence on time has been removed.

Suppose that the mechanical system consists of a single particle of mass $m$, moving under the action of a potential  $V(\mathbf{r},t)$, where $\mathbf{r}=(x,y,z)$ are the Cartesian coordinates of the particle. In such a case the Hamiltonian reads
$$
H=\frac{|\mathbf{p}|^2}{2m}+V(\mathbf{r},t),
$$
and Hamilton-Jacobi equation (\ref{hj}) takes the form
\begin{equation}
\frac{\partial S}{\partial t}+
\frac{1}{2m}\left|\frac{\partial S}{\partial\mathbf{r}}\right|^2+
V(\mathbf{r},t)=0,
\label{hj_p}
\end{equation}
where
$$
\left|\frac{\partial S}{\partial\mathbf{r}}\right|^2=\left(\frac{\partial S}{\partial x}\right)^2+
\left(\frac{\partial S}{\partial y}\right)^2+\left(\frac{\partial S}{\partial z}\right)^2=
|\boldsymbol{\nabla} S|^2.
$$
Moreover in the case of a conservative field (\ref{hj_p}) becomes
\begin{equation}
|\boldsymbol{\nabla} W|^2=2m[E-V(\mathbf{r})].
\label{hjc_p}
\end{equation}

\subsection{Geometrical solution. The Hamiltonian analogy}

Let us now analyze some geometric aspects of the Hamilton principal function $S$ which are necessary to understand the Hamiltonian analogy. For simplicity, we will consider that our system consists of a single particle so that $S$ is given by (\ref{solucio_hjc}):
\begin{equation}
S(\mathbf{r},t)=W(\mathbf{r})-Et,
\label{solucio_hjc2}
\end{equation}
and $W(\mathbf{r})$ satisfies (\ref{hjc_p}). Let us remember that the geometric representation of an equation of the form $f(\mathbf{r})=\ $constant, where $f$ is an arbitrary function, is a surface in the ordinary three-dimensional space. Accordingly, a relation of the form
$$
S(\mathbf{r},t)=\ {\rm constant},
$$
in which time is involved, represents a moving surface. We will detail the time evolution of such a surface when S(r,t) is the Hamilton principal function (\ref{solucio_hjc2}). In what follows, we will not write explicitly the dependence of $S$ on the constants $\boldsymbol{\alpha}$, since we are interested on geometric aspects which do not depend on $\boldsymbol{\alpha}$.

Note that at some arbitrary instant of time $t_0$, the surface $S(\mathbf{r},t_0)=C$, where $C$ is a given constant,  coincides with the surface $W(\mathbf{r})=C+Et_0$ (see Fig. \ref{fig1}). At a later instant $t_0+dt$ the surface $S(\mathbf{r},t_0+dt)=C$ will correspond with the surface  $W(\mathbf{r})=C+E(t_0+dt)$. Therefore, during the interval $dt$ the surface $S=C$ has moved from  $W(\mathbf{r})=C+Et_0$ to $W(\mathbf{r})=C+E(t_0+dt)$. In optics, this is exactly the way wave fronts move. In other words, from a geometrical point of view, {\it the surfaces $S(\mathbf{r},t)=\ {\rm constant}$ can be considered as wave fronts moving in ordinary space}.

\begin{figure}[htbp]
\begin{pspicture}(7,6)
\rput(6.1,4.1){ $W(\mathbf{r})=c$}
\rput(9.1,5.2){ $W(\mathbf{r})=c+Edt$}
\rput(8.15,3.15){ $d\sigma$}
\centerline{\includegraphics[width=6cm]{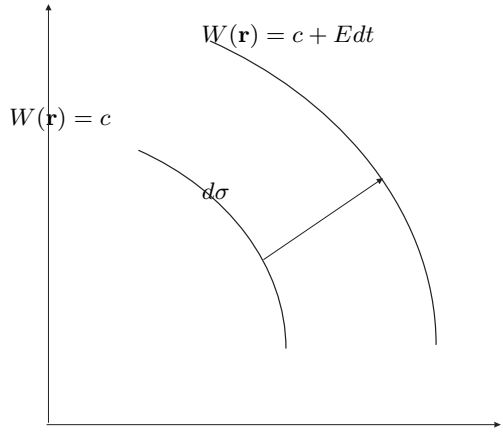}}
\end{pspicture}\vspace*{-2mm}
\caption{Bidimensional representation of the time evolution of the surfaces of constant $S$ from $t_0$ and $t_0+dt_0$ (in the figure we assume $t_0=0$). $d\sigma$ is the distance between two surfaces separated a time interval $dt$. }
\label{fig1}\vspace*{-3mm}
\end{figure}

We will give next a short interlude in optics where this analogy becomes even clearer. Before this, we will derive the velocity of displacement --the ``wave velocity''-- of the surfaces of constant $S$. Let $d\sigma$ be the orthogonal distance between two surfaces separated some small time interval $dt$ 
(see Fig. \ref{fig1}). Then, the displacement velocity is
\begin{equation}
v=\frac{d\sigma}{dt}.
\label{v0}
\end{equation}
On the other hand, if we differentiate the relation $S(\mathbf{r},t)=W(\mathbf{r})-Et=C$, we get $dW=Edt$, or equivalently
$$
\boldsymbol{\nabla}W\cdot d\mathbf{r}=Edt,
$$
but $\boldsymbol{\nabla}W\cdot d\mathbf{r}=|\boldsymbol{\nabla}W|d\sigma$. Hence,
\begin{equation}
v=\frac{E}{|\boldsymbol{\nabla}W|},
\label{v}
\end{equation}
where the value of the gradient of W is calculated through the Hamilton-Jacobi equation. In the case of a particle moving in a conservative field $|\boldsymbol{\nabla}W|$ is given by Equation (\ref{hjc_p}) and 
\begin{equation}
v=\frac{E}{\sqrt{2m[E-V(\mathbf{r})]}},
\label{v2}
\end{equation}
showing that the ``wave velocity'' $v$ is not generally uniform but depends on $\mathbf{r}$.

Let us write (\ref{v}) in a more suggestive way. We know that in terms of $S$, momenta are given by $\mathbf{p}=\partial S/\partial\mathbf{q}$ (see (\ref{solucio_hj})). For a conservative system, $\partial S/\partial\mathbf{q}=\partial W/\partial\mathbf{q}=\boldsymbol{\nabla}W$ and
$$
\mathbf{p}=\boldsymbol{\nabla}W.
$$
Therefore,
\begin{equation}
v=\frac{E}{|\mathbf{p}|}.
\label{v3}
\end{equation}
On the other hand, the gradient $\boldsymbol{\nabla}S=\boldsymbol{\nabla}W$ is a normal vector to the surface $S=$constant and, as $\mathbf{p}$ has the direction of the trajectory, we can conclude that {\it the trajectories of the particle are perpendicular to the surfaces of constant $S$}. As we can see, we can obtain the trajectories by drawing the normal curves to the surfaces $S=$\ constant. Therefore, {\it the mechanical trajectories behave as the light rays of optics}, because the latter are perpendicular to wave fronts and the same happens with particle trajectories which are perpendicular to the ``mechanical wave fronts'' $S=$ constant. This is essentially the Hamiltonian analogy between geometrical optics and classical mechanics.

\section{An interlude in optics}

As the title of this work indicates, the passage from classical to quantum mechanics is carried out using optics as a guide. As we have just seen the Hamiltonian analogy, the starting point of the journey, is precisely the resemblance of mechanical trajectories with light rays of geometrical optics. To proceed further we have to complete the analogy and  establish the association between mechanics and wave optics. In this section we summarize the main aspects of optics that are relevant for the establishment of quantum mechanics. The reader familiar with optics may skip this section.

Let $\phi$ be an electromagnetic perturbation, that is, $\phi$ is any component of the electric field $\mathbf{E}$, or the magnetic field $\mathbf{B}$, or the potential vector $\mathbf{A}$. In certain circumstances as, for instance, in the vacuum in absence of charges or in an electrically neutral medium, Maxwell equations show that the perturbation $\phi$ satisfies the {wave equation}
\begin{equation}
\frac{\partial^2\phi}{\partial t^2}=v^2\nabla^2\phi,
\label{eq_ones}
\end{equation}
where $\nabla^2=\boldsymbol{\nabla}\cdot\boldsymbol{\nabla}$ is the Laplacian operator and $v$ is the velocity of the wave, also known as {phase velocity}. 

The wave equation may have countless solutions depending on the properties of the medium, the dependence of the wave velocity on position and time, the boundary conditions, etc. For stationary and uniform media the wave velocity $v$ is a constant independent of $t$ and $\mathbf{r}$. In this case there exists a simple and special kind of solution which is particularly useful for the development of optics. It is the {monochromatic} or {harmonic plane wave}:   
\begin{equation}
\phi(\mathbf{r},t)=A\exp\left\{i\left[\omega\left(\frac{\mathbf{u}\cdot\mathbf{r}}{v}-t\right)+\delta\right]\right\},
\label{ona_harmonica}
\end{equation}
where $A$, $\omega$ and $\delta$ are constants, $\mathbf{u}$ is a unit vector defining the direction along which the wave propagates. The argument of the imaginary exponential, $\omega(\mathbf{u}\cdot\mathbf{r}/v-t)+\delta$, is an angle called the phase of the wave. Note that on the moving planes,
$$
\mathbf{u}\cdot\mathbf{r}-vt=\ {\rm constant},
$$
where all points have the same phase, the perturbation $\phi(\mathbf{r},t)$ is constant. These plane surfaces, which move perpendicularly to $\mathbf{u}$ with velocity $v$, are called {wave fronts}. 

One interesting feature of the solution (\ref{ona_harmonica}) is that clearly shows the double periodicity in time and space. Indeed $\phi(\mathbf{r},t)$ remains unaltered when $t$ is changed by $ t+T$ and $\mathbf{u}\cdot\mathbf{r}$ by $\mathbf{u}\cdot\mathbf{r}+\lambda$, where $T=2 \pi/\omega$ is the period of the wave, $\nu=1/T$ its frequency and
$\lambda=vT$ the wavelenght. The refractive index is defined by
$$
n=\frac{c}{v}
$$
where $c$ is the speed of light in the vacuum. The wave number $k$ is defined by
$$  
k=\frac{2\pi}{\lambda}=n\frac{2\pi}{\lambda_0}=nk_0,
$$
where $\lambda_0=n\lambda$ and $k_0=k/n$ are respectively the reduced wavelength and the wave number in vacuum. The vector $\mathbf{k}=k\mathbf{u}$ is called the {wave vector} and it has the direction of propagation of the wave, which is orthogonal to the wave fronts. Finally, in terms of the wave vector the monochromatic wave (\ref{ona_harmonica}) can be written as
\begin{equation}
\phi(\mathbf{r},t)=A\exp\left\{i\left(\mathbf{k}\cdot\mathbf{r}-\omega t\right)\right\},
\label{ona1}
\end{equation}
where we have included the factor $e^{i\delta}$ in the amplitude $A$. In an equivalent way, we may write
\begin{equation}
\phi(\mathbf{r},t)=A\exp\left\{i k_0\left(n\mathbf{u}\cdot\mathbf{r}-ct\right)\right\}.
\label{ona2}
\end{equation} 

In spite of its simplicity and usefulness, the monochromatic wave is an idealization rarely found in nature. Real light signals are aggregates of monochromatic waves, the so-called wave trains. Moreover, many media may be {dispersive}  which means that the refractive index varies with the frequency. In these circumstances real light signal take the form of a ``wave packet'' (see Appendix A for a more complete discussion on this issue). The velocity of displacement of a wave packet is called {group velocity}, $v_g$, and in Appendix A we show that it is given by
\begin{equation}
v_g=\frac{d\omega}{dk}.
\label{v_g0}
\end{equation}
In a non dispersive medium frequency and wave number are related through a linear relation, $\omega=kv$, then $v_g=v$ and the group velocity coincides with the phase velocity. 

\subsection{The eikonal equation}

Inhomogeneous media, where the phase velocity and the refractive index depend on the position, constitute another situation often encountered in nature. Now (\ref{ona1}) and (\ref{ona2}) are no longer solutions of the wave equation (\ref{eq_ones}). Nevertheless, one still looks for solutions having the double periodicity in time and space showed by the monochromatic wave. A relatively simple way of getting this consists in changing the constant amplitude, $A$, and the linear term, $n\mathbf{u}\cdot\mathbf{r}$, of Equation (\ref{ona2}), by arbitrary functions of the position. Then, in an inhomogeneous medium we look for solutions of the wave equation in the form
\begin{equation}
\phi(\mathbf{r},t)=A(\mathbf{r})\exp\{ik_0[\mathcal{L}(\mathbf{r})-ct]\},
\label{eikonal}
\end{equation}
where $A(\mathbf{r})$ and $\mathcal{L}(\mathbf{r})$ are real functions. We will see next the conditions these functions must satisfy for Equation (\ref{eikonal}) to be a solution of the wave equation. 

Let us note first that wave fronts are now given by
\begin{equation}
\mathcal{L}(\mathbf{r})-ct=\mbox{ constant},
\label{fronts_ona}
\end{equation}
which represent arbitrary moving surfaces depending on the form of $\mathcal{L}(\mathbf{r})$. We also observe the analogy between these wave fronts and the surfaces $S=$constant of the Hamilton-Jacobi theory:
\begin{equation}
W(\mathbf{r})-Et=\mbox{ constant}.
\label{S_const}
\end{equation}
This ratifies our previous statement obtained by geometric considerations: the surfaces of constant $S$ are like  wave fronts.

The functions $A(\mathbf{r})$ and $\mathcal{L}(\mathbf{r})$ defining the inhomogeneous wave (\ref{eikonal}) are arbitrary and they need to be known. However, the perturbation $\phi(\mathbf{r},t)$ given by (\ref{eikonal}) must obey the wave equation (\ref{eq_ones}). Plugging the former into the latter and separating real and imaginary parts, we get
\begin{equation}
4\pi^2\left[|\boldsymbol{\nabla}\mathcal{L}|^2-n^2\right]-\lambda_0^2\frac{\nabla^2 A}{A}=0,
\label{eq_L}
\end{equation}
\begin{equation}
\nabla^2\mathcal{L}+2\frac{\boldsymbol{\nabla}A}{A}\cdot\boldsymbol{\nabla}\mathcal{L}=0.
\label{eq_A}
\end{equation}
These equations which determine both $A(\mathbf{r})$ and $\mathcal{L}(\mathbf{r})$ also ensure that the expression given in (\ref{eikonal}) is a solution of the wave equation in an inhomogeneous medium where the phase velocity and, consequently, the refractive depend on the position.

Knowing $A(\mathbf{r})$ and $\mathcal{L}(\mathbf{r})$ through the exact solution of (\ref{eq_L}) and (\ref{eq_A}) is rather involved, not to say impossible. The search for approximated solutions is therefore necessary. The help comes from the following empirical fact: the electromagnetic field associated to visible light has short wavelengths, between $4 \times 10^{-5}$ cm and $7 \times 10^{-5}$ cm. Hence, one can expect a good approximation to the laws of light propagation by neglecting the wavelength. In fact, this procedure has proved to be quite appropriate  because phenomena related to non negligible wave lengths, such as diffraction and interference, are apparent only after high precision experiments are performed. Geometrical optics is a branch of optics characterized by neglecting the wavelength. Within this approximation the laws of optics are expressed in the language of geometry and the energy is conveyed by means of definite curves called light rays.
 
We will address the problem posed by (\ref{eq_L}) and (\ref{eq_A}) using the methods of geometrical optics. We then suppose that the wavelength, $\lambda_0$, is small. To be more precise, we assume that $\lambda_0$ is much smaller than the length over which the refractive index substantially changes, that is,
$$
\lambda_0|{\nabla}n|\ll 1.
$$
Taking the limit $\lambda_0\rightarrow 0$ in Equation (\ref{eq_L}) we obtain
\begin{equation}
|{\nabla}\mathcal{L}|^2=n^2.
\label{eq_eikonal}
\end{equation}
The function $\mathcal{L}(\mathbf{r})$ is called the {eikonal} (from Greek $\epsilon\iota\kappa\omega\nu=$ image) and (\ref{eq_eikonal}) is the {eikonal equation}, the fundamental equation of the geometrical optics. Once we know $\mathcal{L}(\mathbf{r})$, the surfaces 
$$
\mathcal{L}(\mathbf{r})=\ {\rm constant}
$$
define wave fronts and light rays are the curves orthogonal to them.

\subsection{The principle of Fermat}

The orthogonality between rays and wave fronts, the so-called ``ray property'' is far from being trivial. In more than two dimensions an arbitrary continuous family of curves cannot be considered as the orthogonal trajectories of some family of surfaces. In mechanics the orthogonality of trajectories to the surfaces $S$= constant, where $S$ is the Hamilton's principal function, holds only because mechanical paths are derivable from a variational principle. Without the principle of least action, the ray property of mechanical paths could not be established.

In optics, the situation is completely analogous and the ray property is a consequence of a variational principle, the {principle of Fermat} (1657). Consider a ray of light propagating in some medium with refractive index $n$. During the time interval $dt$, and due to the property of orthogonality, the ray has moved perpendicularly between two wave fronts  separated $d\sigma$ (see Fig. \ref{fig1}) and has, therefore, traveled the smallest distance between successive fronts. Consequently the time taken to cover this distance is minimum. Any other path would result in greater lengths and longer times. This is Fermat's principle of the ``quickest path'': 

{\it The trajectory of light satisfies the property that if a ray of light travels between two given points of the space, it does so in the least possible time}.

Taking into account that 
$$
dt=\frac{d\sigma}{v}=\frac{n}{c}d\sigma,
$$
where $n$ is the refractive index, we see that minimizing $t$ means minimizing the integral
$$
I=\int_{t_1}^{t_2}\frac{n}{c}d\sigma,
$$
and, since $c$ is constant, we arrive at the principle of Fermat:
\begin{equation}
\delta\left(\int_{t_1}^{t_2}nd\sigma\right)=0. 
\label{fermat}
\end{equation}

The analogy between Fermat's principle (\ref{fermat}) and principle of least action (\ref{pma}) is even clearer when we write the latter in the form called {Maupertuis' principle}, which is one way of writing the variational principle for a particle moving in a conservative field. In this case, $L=T-V$ and from conservation of energy we write $L=2T-E$, where $E$ the energy of the particle. Since $E$ is constant, the variation of the action integral $\int Ldt$ is equal to the variation of the integral $2\int Tdt$. On the other hand, the motion of the particle is $p=mu$ where $u=dl/dt$ is its velocity ($dl$ is the length element of the trajectory of the particle). Thus, $T=p^2/2m=(1/2)p(dl/dt)$ and the integral to be minimized is $2\int Tdt=\int pdl$. Collecting results we arrive at the principle of Maupertuis principle:
\begin{equation}
\delta\left(\int_{t_1}^{t_2}pdl\right)=0.
\label{maupertuis}
\end{equation}
The resemblance with Fermat's principle (\ref{fermat}) is eloquent. The principle (\ref{maupertuis}), a special case of Hamilton's principle (\ref{pma}), was proposed in 1744 by Maupertuis (1698-1759) who enunciated the universal hypothesis that in every event of nature there is some magnitude called ``action'', which is minimal.

After this digression around the world of optics, we are in a better position to go further than the Hamiltonian analogy and make a great leap forward toward new physics.

\section{Wave mechanics}
	
The first step is to realize that the eikonal equation, 
$$
|{\nabla}\mathcal{L}|^2=n^2,
$$
has the same form than the Hamilton-Jacobi equation,
$$
|\boldsymbol{\nabla} W|^2=2m[E-V(\mathbf{r})],
$$
of a particle of mass $m$ moving in a potential field $V(\mathbf{r})$. Looking at both equations we see that Hamilton's characteristic function $W$ plays the same role than the eikonal $\mathcal{L}$, and that the expression $\sqrt{2m(E-V)}$ plays the role of the index of refraction. 

This comparison suggests that classical mechanics might be viewed as the geometrical optics --{\it i.e.}, the limit for small wavelengths-- of an ``undulatory motion'' or ``wave mechanics'' where rays orthogonal to wave fronts would correspond to mechanical paths since the latter enjoy the ray property. This also explains the analogy between Hamilton's and Fermat's principles. Indeed, if we compare the variational principles given by (\ref{fermat}) and (\ref{maupertuis}) we see that the "mechanical refractive index"
$$
n=\sqrt{2m(E-V)}
$$
should be equal or proportional to the particle momentum $p$. This is certainly the case, because from the conservation of the energy 
$$
\frac{p^2}{2m}+V=E,
$$
we see that 
$$
p=\sqrt{2m(E-V)},
$$
which completes the analogy between Hamilton's and Fermat's principles.

Accepting that classical mechanics is the limit for small wavelengths of a hypothetical wave mechanics, two questions arise at once: (i) which are the wavelengths associated to the wave motion? and (ii) what is the ``mechanical wave equation'' approaching the Hamilton-Jacobi equation for vanishing wavelengths?. We will answer the first question in this section and postpone  the second for the next.

Let us note first that mechanics, viewed as a sort of geometrical optics, cannot account for events depending on the wavelength such as interferences and diffractions. We can, nonetheless, speculate about the form of a wave equation that would result in the Hamilton-Jacobi equation as a limit case. However, the correspondence between classical mechanics and wave mechanics is not biunivocal and demands conjectures, in the same manner as geometrical optics is contained in wave optics but not vice versa. 

Having these considerations in mind, let us try to answer the first question posed above. If Hamilton`s characteristic function $W$ is the analogue of the eikonal $\mathcal{L}$ then, recalling that Hamilton's principal function $S$ is related to $W$ by
$$
S=W-Et,
$$
and taking into account that the phase of an optical wave is proportional to $\mathcal{L}(\mathbf{r})-ct$ (cf. (\ref{eikonal})), we make, following Schr\"{o}dinger, the fundamental hypothesis that {\it $S$ is analogous to the phase of the mechanical wave}. Specifically, we assume that Hamilton's principal function is proportional to the phase of the wave:
\begin{equation}
S=K\ {\rm phase},
\label{hipotesi}
\end{equation}
where $K$ is an arbitrary constant to be identified later. Moreover, looking at Equation (\ref{eikonal}), we write the phase of the wave as:
\begin{equation}
\mbox{phase }=k(\mathcal{L}-vt)=2\pi\left(\frac{\mathcal{L}}{\lambda}-\nu t\right),
\label{fase}
\end{equation}
where $k=2\pi/\lambda$ is the wave number, $\lambda$ the wavelength, $\nu=v/\lambda$ the frequency and $v$ the wave velocity. Hence, Equation (\ref{hipotesi}) yields
$$
W(\mathbf{r})-Et=2\pi K\left[\frac{\mathcal{L}(\mathbf{r})}{\lambda}-\nu t\right].
$$
Identifying the spatial part of this identity we obtain the eikonal of the mechanical wave
\begin{equation}
\mathcal{L}(\mathbf{r})=\frac{\lambda}{2\pi K}W(\mathbf{r}).
\label{eikonal2}
\end{equation}
The identification of terms involving time leads to the following relation between energy and frequency
\begin{equation}
E=2\pi K\nu.
\label{planck0}
\end{equation}
This equation coincides with Planck's quantum hypothesis, $E=h\nu$, if we recognized the arbitrary constant $K$, introduced in (\ref{hipotesi}), with Planck's reduced constant
\begin{equation}
K=\frac{h}{2\pi}=\hbar.
\label{K}
\end{equation}

Having determined the eikonal and the frequency we can easily obtain the wavelength, answering thus our first question. We know that $\lambda= v/\nu,$ where $v$ is the wave velocity and $\nu$ the frequency. Moreover, from Equation (\ref{v3}) we have $v=E/p,$ where $p$ is the momentum of the particle and $E$ its energy. On the other hand, by combining (\ref{planck0}) and (\ref{K}) we see that $\nu=E/h$. Collecting results we arrive at,
\begin{equation}
\lambda=\frac{h}{p},
\label{de_Broglie}
\end{equation}
which is the celebrated de Broglie's wavelength.

One may argue, following the historical course of events, that the wavelength (\ref{de_Broglie}) was proposed by Louis de Broglie in 1924, two years before the birth of wave mechanics occurred in 1926. Therefore, de Broglie could not have obtained his wavelength in the way just explained which is based on Schr\"{o}dinger's assumption (\ref{hipotesi}). 

Indeed, the dual nature of light was conjectured by Einstein in 1905. Light propagates as an electromagnetic wave but it interacts with matter as if its energy were concentrated in ``packets'', the {\it quanta} of the energy. This conjecture was experimentally verified in 1916 by Millikan's high precision measurements on the photoelectric effect. The dual conception of material particles is due to Louis de Broglie who argued, in 1924, that if any wavelength had to be associated to a particle with momentum $p$ in an invariant relativistic way, this could only be achieved by Equation (\ref{de_Broglie}). In Appendix B we give an outline of how (\ref{de_Broglie}) was originally derived by de Broglie. The work of de Broglie and the Hamiltonian analogy are fundamental milestones in the development of quantum mechanics.

Summarizing what we have seen up to this point we can say that a wave mechanics, which for small wavelengths coincide with classical mechanics, would associate to each particle of energy $E$ and momentum $p$ a frequency given by Planck's law:
$$
\nu=\frac{E}{h},
$$
and a wavelength given by the de Broglie's expression (\ref{de_Broglie}). Observe that we have arrived at this conclusion only by assuming that Hamilton's principal function is proportional to the phase of the mechanical wave and using Planck's quantum hypothesis in order to identify the constant of proportionality  with Planck's reduced constant $\hbar$.

\section{The stationary Schr\"{o}dinger equation}

We are now ready to answer the second question formulated at the beginning of the last section: what is the wave equation that for short wavelengths goes to Hamilton-Jacobi equation?

Using (\ref{hipotesi}) and (\ref{K}) we write the phase of the mechanical wave as
\begin{equation}
\mbox{phase }=\frac{S}{\hbar}=\frac{W(\mathbf{r})-Et}{\hbar}.
\label{fase2}
\end{equation}
Suppose, in addition, that it exists a wave field such that its intensity indicates the density of particles, in the same way that the intensity of the electromagnetic field denotes the density of photons. Furthermore, let us assume that it is a scalar field given by a single function $\Psi(\mathbf{r},t)$ which, according to its supposed wave  properties, we write in the form
\begin{equation}
\Psi(\mathbf{r},t)=\Psi_0(\mathbf{r})e^{iS(\mathbf{r},t)/\hbar}.
\label{f_ona1}
\end{equation}
Applying equation (\ref{fase2}) we can separate the spatial and temporal dependence of the field and write
\begin{equation}
\Psi(\mathbf{r},t)=\psi(\mathbf{r})e^{-iEt/\hbar},
\label{f_ona2}
\end{equation}
where
\begin{equation}
\psi(\mathbf{r})=\Psi_0(\mathbf{r})e^{iW(\mathbf{r})/\hbar}.
\label{f_ona3}
\end{equation}

Let us now impose that $\Psi(\mathbf{r},t)$ obeys the wave equation
\begin{equation}
\frac{\partial^2\Psi}{\partial t^2}=v^2\nabla^2\Psi.
\label{eq_ones2}
\end{equation}
Substituting the expression for $\Psi(\mathbf{r},t)$ given by (\ref{f_ona2}) we see  that $\psi(\mathbf{r})$ satisfies the equation
\begin{equation}
v^2\nabla^2\psi+\frac{E^2}{\hbar^2}\psi=0.
\label{ES0}
\end{equation}
Replacing into this equation the expression (\ref{v2}) of the velocity of the wave associated to a particle moving under a potential $V(\mathbf{r})$, 
$$
v=\frac{E}{\sqrt{2m(E-V)}},
$$
we see that $\psi(\mathbf{r})$ satisfies 
\begin{equation}
\frac{\hbar^2}{2m}\nabla^2\psi+(E-V)\psi=0.
\label{ES}
\end{equation}

This is the {Schr\"{o}dinger wave equation} for a free particle in a scalar potential field. As it is a time-independent equation, we can interpret it as describing the stationary, e.g. periodic, motion of a particle in a field of force. But we could likewise apply it to the stationary beams formed by many particles, one after another and under identical conditions, as they appear, for instance, in electron optics. Whatever may be the case, it seems reasonable to suppose, according to {\it Born's interpretation}, that the intensity of the mechanical wave, which is proportional to the square of the absolute value of the amplitude  
$$
|\Psi|^2=\Psi\Psi^*,
$$
is proportional to the particle density at the point $\mathbf{r}=(x,y,z)$ measured over long times; or, what is the same in this case, that is proportional to the probability that the particle will be observed at this spot at any instant. More precisely: 

\medskip

$|\Psi|^2dxdydz$ {\it is proportional to the probability that a particle is observed inside some element of volume $dxdydz$ centered at the point $\mathbf{r}=(x,y,z)$}. 

\medskip

Let us note, however, that the notion of waves which do not transmit energy or momentum, but which determine probability, had become familiar to theoretical physicists from the (unsuccessful) Bohr-Karmers-Slater theory of 1924 \cite{whitt}.

We will finally mention that the potential function $V(\mathbf{r})$ which occurs in the Schr\"{o}dinger equation can posses singularities at certain points or at infinity. These points (and eventually with others) will be singular points of the differential equation (\ref{ES}). Since $\psi(\mathbf{r})$ is to represent a wave in the stationary state, it must be free of singularities; and therefore Schr\"{o}dinger laid down the condition that {\it the solution of the wave equation corresponding to a stationary state must be one-valued, finite and free from singularities}, even at the  singularities of $V(\mathbf{r})$. 

As we know Equation (\ref{ES}) possesses solutions of this kind only for certain special values of the constant $E$, known as {eigenvalues}. These eigenvalues will be the only values of the energy that the particle can have when it is in a stationary state. No wonder that the title Schr\"{o}dinger gave to his first paper was precisely ``Quantification as a problem of eigenvalues''. 

\section{Generalizations of the Schr\"{o}dinger equation}

Equation (\ref{ES}) is valid for a single particle moving under a time-independent ({\it i.e.} conservative) force field. But, what is the Schr\"{o}dinger equation for an arbitrary mechanical system, for example a system of particles, which in the most general case can be non-conservative?

Let us therefore consider a non-conservative system with $n$ degrees of freedom with a Hamiltonian of the form
\begin{equation}
H(\mathbf{q},\mathbf{p},t)=\sum_{k,l}b_{kl}(\mathbf{q},t)p_kp_l+
V(\mathbf{q},t),
\label{hnc}
\end{equation}
where $b_{kl}$ ($k,l=1,2,\cdots,n$) is a positive definite matrix. In this case the Hamilton-Jacobi equation (\ref{hj}) reads 
\begin{equation}
\frac{\partial S}{\partial t}+
\sum_{k,l}b_{kl}(\mathbf{q},t)\frac{\partial S}{\partial q_k}\frac{\partial S}{\partial q_l}+V(\mathbf{q},t)=0.
\label{hj_gen}
\end{equation}

Following Schr\"{o}dinger's hypothesis (\ref{hipotesi}), we suppose that the phase of the mechanical wave is proportional to Hamilton's principal function. Consequently, we assume that the wave function has the form 
\begin{equation}
\Psi(\mathbf{q},t)=\exp\left\{A(\mathbf{q},t)+
iS(\mathbf{q},t)/\hbar\right\},
\label{f_ona_gen}
\end{equation}
where the amplitude is given by 
$$
\Psi_0(\mathbf{q},t)=e^{A(\mathbf{q},t)}.
$$
Differentiating (\ref{f_ona_gen}) twice we get
\begin{equation}
\frac{\partial^2 \Psi}{\partial q_k\partial q_l}=
\Biggl\{\frac{\partial^2 A}{\partial q_k\partial q_l}+
\frac{\partial A}{\partial q_k}\frac{\partial A}{\partial q_l}+ 
\frac{i}{\hbar}\Biggl[
\frac{\partial^2 S}{\partial q_k\partial q_l} 
+\frac{\partial A}{\partial q_k}\frac{\partial S}{\partial q_l}+
\frac{\partial A}{\partial q_l}\frac{\partial S}{\partial q_k}\Biggr]
- \frac{1}{\hbar^2}
\frac{\partial S}{\partial q_k}\frac{\partial S}{\partial q_l}\Biggr\}
\Psi.
\label{lap_Psi}
\end{equation}
If we take into account that $\hbar$ is very small ($\hbar=1.055 \times 10^{-34} Js$) then all terms of the second member of (\ref{lap_Psi}) are very small compared to the last term. Hence,
\begin{equation}
\frac{\partial^2 \Psi}{\partial q_k\partial q_l}\simeq 
-\frac{1}{\hbar^2}
\left(\frac{\partial S}{\partial q_k}\frac{\partial S}{\partial q_l}\right)\Psi.
\label{lap_aprox2}
\end{equation}

On the other hand, the time derivative of the wave function (\ref{f_ona_gen}) is
$$
\frac{\partial \Psi}{\partial t}=
\left[\frac{\partial A}{\partial t}+\frac{i}{\hbar}
\frac{\partial S}{\partial t}\right]\Psi, 
$$
which, considering again that $\hbar$ is very small, reads
\begin{equation}
\frac{\partial \Psi}{\partial t}\simeq \frac{i}{\hbar}
\frac{\partial S}{\partial t}\Psi.
\label{time_derivative}
\end{equation}
Substituting into the Hamilton-Jacobi equation (\ref{hj_gen}) the expressions of $(\partial S/\partial q_k)(\partial S/\partial q_l)$ and $\partial S/\partial t$ given respectively by (\ref{lap_aprox2}) and (\ref{time_derivative}), we see that $\Psi(\mathbf{q},t)$ satisfies the equation
\begin{equation}
\sum_{k,l}b_{kl}(\mathbf{q},t)
\frac{\partial^2 \Psi}{\partial q_k\partial q_l}-
\frac{1}{\hbar^2}V(\mathbf{q},t)\Psi+
\frac{i}{\hbar}\frac{\partial \Psi}{\partial t}=0,
\label{ES_gen}
\end{equation}
which is the Schr\"{o}dinger equation for this general case.

Before proceeding further let us recall that classical mechanics and, hence, Hamilton-Jacobi equation (\ref{hj_gen}) are obtained from wave mechanics in the limit of vanishing wavelengths. But from the de Broglie's relation, $\lambda=h/p$, we see that the limit $\lambda\rightarrow 0$ is equivalent to $h\rightarrow 0$. A fact that sustains the
above approximations which involve the neglect of powers of $\hbar$. One is, however, left with the impression that the foregoing derivation connecting Hamilton-Jacobi and Schr\"{o}dinger equations is not exact, but only approximate. This conclusion is, nonetheless, erroneous: it can be shown that both equations are rigorously equivalent (see Ref.  \cite{whitt} for more details). 

We will next write the Schr\"{o}dinger equation (\ref{ES_gen}) in a more compact and suggestive way. We first rewrite  (\ref{ES_gen}) as
$$
-\hbar^2\sum_{k,l}b_{kl}(\mathbf{q},t)
\frac{\partial^2 \Psi}{\partial q_k\partial q_l}+V(\mathbf{q},t)\Psi=
i\hbar\frac{\partial \Psi}{\partial t},
$$
or, in the equivalent way,
\begin{equation}
\Biggl[\sum_{k,l}b_{kl}(\mathbf{q},t)
\left(\frac{\hbar}{i}\frac{\partial}{\partial q_k}\right)
\left(\frac{\hbar}{i}\frac{\partial}{\partial q_l}\right)+
V(\mathbf{q},t)\Biggr]\Psi=i\hbar\frac{\partial \Psi}{\partial t}.
\label{ES_gen_bis}
\end{equation}
But from the expression of the Hamiltonian given by (\ref{hnc}) we can write in a formal way
\begin{equation}
\sum_{k,l}b_{kl}(\mathbf{q},t)
\left(\frac{\hbar}{i}\frac{\partial}{\partial q_k}\right)
\left(\frac{\hbar}{i}\frac{\partial}{\partial q_l}\right)+
V(\mathbf{q},t)=H\left(\mathbf{q},\frac{\hbar}{i}\frac{\partial}{\partial\mathbf{q}},t\right).
\end{equation}
Therefore (\ref{ES_gen_bis}) is written as
\begin{equation}
i\hbar\frac{\partial \Psi}{\partial t}=
H\left(\mathbf{q},\frac{\hbar}{i}\frac{\partial}{\partial\mathbf{q}},t\right)\Psi,
\label{ES_gen2}
\end{equation}
which is the {general Schr\"{o}dinger equation} valid for any mechanical system with a general non-conservative Hamiltonian of the form given by (\ref{hnc}). 

\medskip

One relevant remark should be made at this point since the generalization of the Schr\"{o}dinger equation given by 
(\ref{ES_gen2}) provides a link between wave mechanics and matrix mechanics. The latter, developed in 1925 by Heissenberg, Born, Jordan and Dirac, is based on abstract formulations in terms of operators. The  basic relations are the commutation rules, which in the case of systems of one degree of freedom reduce to
\begin{equation}
qp-pq=i\hbar,
\label{comm}
\end{equation}
where $q$ and $p$ are operators corresponding to the conjugated position and momentum. 

Now, looking at the general Schr\"{o}dinger equation (\ref{ES_gen2}), we observe that we could have obtained that equation had we made the replacements
\begin{equation}
\mathbf{p} \longrightarrow \frac{\hbar}{i}\frac{\partial}{\partial\mathbf{q}},\qquad\qquad 
E \longrightarrow -\frac{\hbar}{i}\frac{\partial}
{\partial t}
\label{link}
\end{equation}
in the equation of the energy $H(\mathbf{q},\mathbf{p},t)=E$ and thereafter operating on the wave function $\Psi$. Note, however, that $E$ is not constant and the energy equation $H=E$ is purely formal, just a notational ease. 

We thus see that (\ref{link}) provides the aforementioned link between wave and matrix mechanics. Schr\"{o}dinger himself did not, at the outset of his researches, suspect any connection between his theory and matrix mechanics. He, for instance, wrote in 1926: ``I naturally knew about his [Heisenberg's] theory, but was discouraged by what appeared to me as very difficult methods of transcendental algebra'' \cite{whitt}. But little afterward he proved  that the two theories were actually equivalent \cite{schrodinger}. Thus, for instance, the commutation relation (\ref{comm}) becomes an obvious identity if, using (\ref{link}), we write $p=(\hbar/i)\partial/\partial q$. Indeed
$$
q\left(\frac{\hbar}{i}\frac{\partial}{\partial q}\right)\psi-\frac{\hbar}{i}\frac{\partial}{\partial q}(q\psi)=
i\hbar\psi.
$$

\medskip

After this brief excursion on matrix mechanics, we return to the generalized Schr\"{o}dinger equation. For a conservative system (\ref{ES_gen2}) reads
\begin{equation}
i\hbar\frac{\partial \Psi}{\partial t}=
H\left(\mathbf{q},\frac{\hbar}{i}\frac{\partial}{\partial\mathbf{q}}\right)\Psi.
\label{ES_gen_con}
\end{equation}
Now $H=E$ is constant and we may look for solutions representing harmonic waves of angular frequency $\omega=E/\hbar$:
\begin{equation}
\Psi(\mathbf{q},t)=\psi(\mathbf{q})e^{-iEt/\hbar}.
\label{f_ona_harmonica}
\end{equation}
Substituting into (\ref{ES_gen_con}) we see that the wave amplitude $\psi(\mathbf{q})$ obeys the equation
\begin{equation}
H\left(\mathbf{q},\frac{\hbar}{i}\frac{\partial}{\partial\mathbf{q}}\right)\psi=E\psi.
\label{ES_gen_con2}
\end{equation}
That is,
\begin{equation}
\hbar^2\sum_{k,l}b_{kl}(\mathbf{q})
\frac{\partial^2 \psi}{\partial q_k\partial q_l} +
[E-V(\mathbf{q})]\psi=0.
\label{ES_gen_con3}
\end{equation}
Either (\ref{ES_gen_con2}) or (\ref{ES_gen_con3}) constitute the stationary Schr\"{o}dinger equation for systems with $n$ degrees of freedom. 

Finally, if the system consists of a single particle of mass $m$, then 
$$
b_{kl}=\delta_{kl}/2m
$$
($\delta_{kl}$ is the Kronecker symbol) and (\ref{ES_gen_con3}) reduces to the Schr\"{o}dinger equation previously obtained (cf. (\ref{ES})).

\section{Wave packets and classical motion}

Before ending this paper we will see, through a specific example, how the wave function is related to the classical motion of the system. As we have seen (see also Appendix B) classical  motion has to be associated not only to a wave but to a {\it wave packet}, i.e., a group of wave functions. It was Schr\"{o}dinger who in 1926, in a paper entitled ``The Continuous Transition from Micro- to Macro-mechanics'' (reproduced in Ref. \cite{schrodinger}), showed how to construct, for the harmonic oscillator, a wave packet with its amplitude very nearly concentrated at a single point which corresponds to the classical motion of the oscillator.

The Schr\"{o}dinger equation for the unidimensional harmonic oscillator of mass $m$ and frequency $\omega$ is
$$
i\hbar\frac{\partial\Psi}{\partial t}=-\frac{\hbar^2}{2m}\frac{\partial^2\Psi}{\partial q^2}+
\frac{1}{2}m\omega^2q^2\Psi.
$$
As can be seen by direct substitution this differential equation is satisfied by
\begin{equation}
\Psi(q,t)=\exp\Bigl\{-i\omega t/2+C(2m\omega/\hbar)e^{i\omega t}q
-(m\omega/2\hbar)q^2-C^2\left(1-e^{2i\omega t}\right)/2\Bigr\},
\end{equation}
where $C$ is any constant. Splitting the real and imaginary parts we write
\begin{equation}
\Psi(q,t)=A(q,t)e^{-i\varphi(q,t)},
\label{paquet2}
\end{equation}
where
\begin{equation}
A(q,t)=\exp\left\{-m\omega\left(q-
a\cos\omega t\right)^2/2\hbar\right\},
\label{A2}
\end{equation}
\begin{equation}
a=C(2\hbar/m\omega)^{1/2},
\label{a2}
\end{equation}
and
\begin{equation}
\varphi(q,t)=\omega t/2+C\bigl[q(2m\omega/\hbar)^{1/2}
-C\cos\omega t\bigr]\sin\omega t.
\label{varphi}
\end{equation}

Equation (\ref{paquet2}) shows that the wave function $\Psi(q,t)$ represents a wave packet of amplitude $A$ and phase $\varphi$ (see Equation (\ref{phi}) of Appendix A). Moreover the amplitude (\ref{A2}) is a Gaussian centered at the oscillating point:
$$
q=a\cos\omega t.
$$
Therefore, the maximum value of the amplitude of the wave packet moves as a classical oscillator of angular frequency $\omega$ and amplitude $a$. The energy of such an oscillator is  $E=(1/2)ma^2\omega^2$ which, after using (\ref{a2}), reads
$$
E=C^2\hbar\omega.
$$
But, as Schr\"{o}dinger had previously shown by solving the stationary equation, the energy of the quantum harmonic oscillator is 
$$
E=(n+1/2)\hbar\omega
$$ 
($n=0,1,2,\cdots$). This sets the value of $C$ at
\begin{equation}
C^2=n+1/2, 
\label{k_n}
\end{equation}
where $n=0,1,2,\cdots$ is any natural number.

On the other hand, $\Psi\Psi^*$ is the probability density function of finding the oscillator at time $t$ in position q. From (\ref{paquet2}) and (\ref{A2}) we have
\begin{equation}
\Psi\Psi^*=\exp\left\{-\frac{\left(q-a\cos\omega t\right)^2}
{\hbar/m\omega}\right\}.
\label{prob}
\end{equation}
If $\hbar$ is very small (as it is) and $C$ is very large, while $a$ given by (\ref{a2}) is finite, then the probability density, $\Psi\Psi^*$, is negligibly small except when the numerator of the argument of the exponential is approximately zero; that is when 
$$
q=a\cos\omega t,
$$
which is the equation of motion of the classical oscillator. Therefore, the classical equation of motion is, with very high probability, the most likely state of the oscillator. Recall that we have arrived at the classical motion assuming a large value of $C$. However this implies, owing to (\ref{k_n}), a large value for $n$. In other words, we recover the classical dynamics in the limit of high quantum numbers. This is precisely {\it Bohr's correspondence principle}.

\section{Summary}

The conceptual parallelism between light rays and  trajectories of material particles was first noticed at the beginning of the 18th century by Jean Bernouilli (1667-1748), who tried to develop a mechanical theory for the refractive index \cite{lanczos}. More than one century elapsed before William Rowan Hamilton (1805-1865) fully developed this idea realizing that the problems arising in mechanics and geometrical optics could be treated in the same unified way. 

Hamilton worked with ``characteristic'' and ``principal'' functions, both within mechanics and optics. These functions have the property that by simple differentiations we can find particle trajectories and optical paths. Moreover, both in mechanics and optics, the characteristic function satisfies the same type of differential equation, whose solution, with the appropriate boundary conditions, is equivalent to the solution of the equations of motion.

Erwin Schr\"{o}dinger (1887-1961) went beyond the Hamiltonian analogy. Following previous ideas from Louis de Broglie (1892-1987) and Albert Einstein (1879-1955), Schr\"{o}dinger wondered what would be a hypothetical wave motion having classical mechanics as a limiting case. This would certainly imply that there is a wave associated to each particle. The characterization of such a wave motion is readily undertaken  by assuming that the phase of the mechanical wave is proportional to Hamilton's principal function. Proceeding along this way, one can easily identify which are the wavelengths and frequencies of the mechanical wave, as well as obtaining the wave equation, which turns out to be the celebrated Schr\"{o}dinger equation. 

Once the merging between classical mechanics and optics was completed, there appeared a new field: wave mechanics. On the other hand, from the equivalence between wave mechanics and matrix mechanics --the latter developed in 1925 by Werner Heisenberg (1901-1976), Max Born (1882-1970), Pasqual Jordan (1902-1980) and P. A. M. Dirac (1902-1984)-- emerged {quantum mechanics}, one of the longest-range scientific revolutions of the 20th century.

\section*{Acknowledgements}

Partial finantial support from Direcci\'{o}n general de Investigaci\'{o}n under Contract No.FIS2006-05204 is acknowledged.

\appendix

\section{Light signals, wave packets and group velocity.}
	
When the refractive index of the medium is independent of the frequency, the phase velocity $v$ is the only speed  appearing in the problem. However, many material media are {dispersive}, that is to say, the index of refraction is a function of the frequency. In these cases $v$ is not unique, for the phase velocity depends now on the frequency of the wave: different colours travel with different speeds.

On the other hand, monochromatic waves are idealizations that cannot be completely realized in practice. Any light signal is a mixture of waves of different frequencies, the so-called wave trains. Even in cases where the signal is supposed to have been emitted by ``monochromatic sources'' (for instance atoms) the corresponding wave train is not a pure harmonic wave but a superposition of many harmonic waves of different frequencies. Therefore, in a dispersive medium, where each component of the signal propagates at different speeds, the concept of phase velocity loses its practical meaning. 

We know, however, that real light signals have finite length, in other words, they fill only finite regions of the space. In this situation a new concept arises: the {signal velocity}, as the speed of propagation of the bulk of the signal. Unfortunately, a light signal, as it progresses through any medium, gets distorted. A fact that hardens the way on how to define its velocity.

Let us address the relatively simple case in which a given signal $\phi$ can be represented by an aggregate of monochromatic waves, all of them traveling along the same direction, and in a dispersive medium characterized by a refractive index, $n=n(\omega)$, depending on the frequency. Assuming without loss of generality that the common direction of propagation is the $x$ axis, we write 
\begin{equation}
\phi(x,t)=
\int_{-\infty}^{\infty} a(\omega)\exp\{i[k(\omega)x-\omega t]\}
d\omega,
\label{grup}
\end{equation}
where $k(\omega)=k(-\omega)$ and $a(-\omega)=a^*(\omega)$. 

By direct substitution is not difficult to see that $\phi(x,t)$ is a solution of the wave equation.  This accounts for the name ``polychromatic waves'' sometimes given to expressions like (\ref{grup}). On the other hand, it also implies that the frequency and the wave number are related by 
$$
k(\omega)=n(\omega)\frac{\omega}{c},
$$
where $n(\omega)$ is the refractive index.  

A case which often appears in practical situations is that of a {quasi-monochromatic} wave. In this case the amplitude $a(\omega)$ of the aggregating waves, only differs from zero on a relatively limited range of frequencies around certain value $\omega_1$, where the amplitude is maximum. This is the case, for instance, when $a(\omega)$ is a narrow Gaussian centered at $\omega_1$. Now the light signal represented by $\phi(x,t)$ is also confined within a finite region and we speak of a ``wave packet'' (see Fig. \ref{fig2}). Moreover, as far as dispersion and absorption by the medium remain small, it is possible to identify the velocity of the signal with the ``group velocity'' which we will define below (see Equation (\ref{v_g})). \footnote{The idea of group velocity was seemingly proposed by Hamilton in 1839. The distinction between phase velocity and group velocity, as well as their relation with the problem of the measurement of the speed of light was Lord Rayleigh's (1877) \cite{bornwolf,brillouin}.} 

\begin{figure}[htbp]
\begin{pspicture}(8.5,4)
\rput(6.5,3.5){\small $\phi(x,t_0)$}
\rput(7.7,1.7){\small $x$}
\rput(10.65,.9){ \small $\omega_1$}
\rput(12,1.2){\small  $\omega$}
\rput(11.2,3){ \small $a(\omega)$}
\centerline{\includegraphics[width=8.5cm]{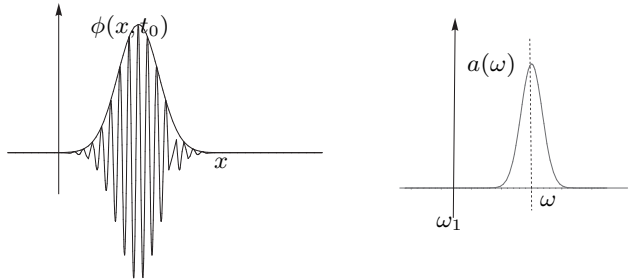}}
\end{pspicture}
\caption{Wave packet at a given instant of time $t_0$ and one of its Fourier components.}
\label{fig2}
\end{figure}

Suppose then a wave packet whose amplitude presents a sharp maximum at a given frequency $\omega_1$ and quickly decays to zero outside this maximum. If, in addition, we assume that the medium is not highly dispersive, then $k(\omega)$ will be a slowly varying function of the frequency $\omega$ for which we can perform the linear approximation in the integral appearing in (\ref{grup}):
\begin{equation}
k(\omega)\simeq k_1+\alpha(\omega-\omega_1)+\cdots,
\label{linear}
\end{equation}
here $k_1=k(\omega_1)$ and $\alpha=dk/d\omega|_1$. Substituting this approximation into (\ref{grup}) and reorganizing terms we get
\begin{equation}
\phi(x,t)\simeq A(x,t)
\exp\{i[k_1x -\omega_1t]\},
\label{phi}
\end{equation}
where 
\begin{equation}
A(x,t)=\int_{-\infty}^{\infty}a(\omega)\exp\{i(\omega-\omega_1)(\alpha x-t)\}d\omega.
\label{A}
\end{equation}

In many cases, $A(x,t)$ is a slowly varying function. In this situations we see from (\ref{phi}) that the group can be  approximately considered as a wave of frequency $\omega_1$ modulated by an amplitude $A$ which is slowly distorted. The traveling speed of the packet, that is, the {group velocity} $v_g$, is the velocity at which the amplitude $A(x,t)$ moves. In other words, $v_g$ is the speed of some observer seeing $A$ stationary. 

Since the amplitude changes very slowly the wave group keeps approximately its shape. Moreover, we see from (\ref{A}) that the dependence of $A$ on the space $x$ and time $t$ is only through the combination $A(\alpha x-t)$. Hence, since $A\approx$ constant, the packet moves with the speed $v_g=1/\alpha$. That is (see (\ref{linear})),
\begin{equation}
v_g=\left.\frac{d\omega}{dk}\right|_1.
\label{v_g}
\end{equation}

The relation between the phase velocity, $v$, and the group velocity, $v_g$, can be obtained by differentiating the equation  $\omega=kv$:
\begin{equation}
v_g=v+k\frac{dv}{dk}.
\label{v_fg}
\end{equation}
Therefore, in a non dispersive medium where the phase velocity is independent of k, $dv/dk=0$, and group and phase velocities coincide.

We finally observe that the amplitude is related to the energy density carried by the wave (the square of the amplitude is proportional to the energy density). Therefore, in the approximation just considered, {\it the energy propagates with the group velocity} because this is the speed at which the amplitude travels. Nevertheless, this is not always true. For example, in the so-called anomalous dispersion regions, the group velocity can overcome the speed of light or even become negative. In these cases the group velocity has no physical meaning.

\section{The de Broglie's wavelength}

The analogy between Fermat's and Hamilton's principles aroused in de Broglie the need for a deeper understanding between corpuscular dynamics and wave propagation. Following this line of reasoning, De Broglie supposed in 1924 that,  with the motion of any material particle, there is associated a wave packet in such a way that {\it the velocity of the particle is equal to the group velocity of the wave packet}. Let $m$ and $u$ be respectively the mass and the velocity of the particle. It is also assumed that the frequency of the associated wave is given by Planck's law 
$$
E=\hbar\omega,
$$
where $E$ the energy of the particle. Then, using the expression for the energy given by the special relativity, we write 
$$
\hbar\omega=\frac{mc^2}{\sqrt{1-u^2/c^2}}.
$$
Under de Broglie's assumption the velocity of the particle $u$ equals the group velocity, (\ref{v_g0}), we thus have 
$$
u=\frac{d\omega}{dk}.
$$
Hence
$$
dk=\frac{d\omega}{u}=\frac{mc^2}
{\hbar u}d\left(1-\frac{u^2}{c^2}\right)^{-1/2}.
$$
That is,
$$
dk=\frac{m du}
{\hbar(1-u^2/c^2)^{3/2}},
$$
which, after recalling the relativistic form of the momentum,
$$
p=\frac{mu}{\sqrt{1-u^2/c^2}},
$$
can be written as
$$
dk=\frac{dp}{\hbar}.
$$
Integrating we have
\begin{equation}
p=\hbar k
\label{p}
\end{equation}
which, after taking into account that $k=2\pi/\lambda$, leads at once to the de Broglie's wavelength (\ref{de_Broglie}):
$$
\lambda=\frac hp.
$$

Finally, the phase velocity of the de Broglie's wave is
$$
v=\lambda\nu=\frac{h}{p}\frac{E}{h}=\frac{E}{p}.
$$
For the relativistic free particle $E/p=c^2/u$, and
$$
v=\frac{c^2}{u},
$$
which is the phase velocity of the wave associated to the particle. Note that for photons (and for other particles whose mass at rest is zero) $v=u=c$.

\end{document}